\documentclass{elsart}
\usepackage{amsmath}
\usepackage{amssymb}
\usepackage{graphicx}
\newtheorem{theorem}{Theorem}
\newtheorem{definition}{Definition}

\begin{document}

\begin{frontmatter}

\title{On Exponential Time Lower Bound of Knapsack under Backtracking}

\author{Xin Li, Tian Liu}
\address{Key laboratory of High Confidence Software Technologies, Ministry of Education}
\address{Software Institute, School of EECS, Peking University, P.R.China}
\address{Email: lixin.pku@gmail.com, lt@pku.edu.cn}

\begin{abstract}
M.Aleknovich et al. have recently proposed a model of algorithms, called BT model, which generalizes both the priority model of Borodin, Nielson and Rackoff, as well as a simple dynamic programming model by Woeginger. BT model can be further divided into three kinds of fixed, adaptive and fully adaptive ones. They have proved exponential time lower bounds of exact and approximation algorithms under adaptive BT model for Knapsack problem. Their exact lower bound is $\Omega(2^{0.5n}/\sqrt{n})$, in this paper, we slightly improve the exact lower bound to about $\Omega(2^{0.69n}/\sqrt{n})$, by the same technique, with related parameters optimized.
\end{abstract}

\end{frontmatter}

\section{Introduction to Backtracking model}

M.Aleknovich et. al have proposed a backtracking model in \cite{1} which generalizes both the priority model of Borodin, Nielson and Rackoff \cite{6}, as well as a simple dynamic programming model due to Woeginger \cite{7}, and hence spans a wide spectrum of algorithms. Borodin then shared his insights into these models in \cite{8}. The definition of backtracking model is as follows.

Let $\mathcal{D}$ be an arbitrary data domain that contains objects $D_i$ called data items. Let $H$ be a set, representing the set of allowable decisions for a data item. For example, for the knapsack problem, a natural choice for $\mathcal{D}$ would be the set of all pairs $(x,p)$ where $x$ is a weight and $p$ is a profit; the natural choice for $H$ is ${0,1}$ where 0 is the decision to reject an item and 1 is the decision to accept an item.

A Backtracking search/optimization problem $P$ is specified by a pair $(\mathcal{D}_P,f_P)$ where $\mathcal{D}_P$ is the underlying domain, and $f_P$ is a family of objective functions, $f_P^n:(D_1,...,D_n,a_1,...,a_n) \mapsto \mathcal{R}$, where $a_1,...,a_n$ is a set of variables that range over $H$, and $D_1,...,D_n$ is a set of variables that range over $\mathcal{D}$. On input $I=D_1,...D_n \in \mathcal{D}$, the goal is to assign each $a_i$ a value in $H$ so as to maximize (or minimize) $f_P^n$. A search problem is a special case where $f_P^n$ outputs either 0 or 1.

For any domain $S$ write $\mathcal{O}(S)$ for the set of all orderings of elements of $S$.
\begin{definition}
A backtracking algorithm $\mathcal{A}$ for problem $P=(\mathcal{D},\{f^n\})$ consists of the ordering functions
\begin{displaymath}
r_\mathcal{A}^k: \mathcal{D}^k \times H^k \mapsto \mathcal{O}(\mathcal{D})
\end{displaymath}
and the choice functions
\begin{displaymath}
c_\mathcal{A}^k: \mathcal{D}^{k+1} \times H^k \mapsto \mathcal{O}(H\cup\{\perp\},
\end{displaymath}
where $k=0,...,n-1$.
\end{definition}

There are three classes of BT algorithms.
\begin{itemize}
\item \emph{Fixed} algorithms: $r_\mathcal{A}^k$ does not depend upon any of its arguments.
\item \emph{Adaptive} algorithms: $r_\mathcal{A}^k$ depends on $D_1,...,D_k$ but not on $a_1,...,a_k$.
\item \emph{Fully adaptive} algorithms: $r_\mathcal{A}^k$ depends on both $D_1,...,D_k$ and $a_1,...,a_k$.
\end{itemize}

The value of $r_\mathcal{A}^k$ specifies the order to consider the remaining items, given that the choices about the first $k$ items have been made; the value of $c_\mathcal{A}^k$ specifies the order to make possible decisions about $D_{k+1}$. For more detailed explanation of BT model, refer to \cite{1}. But it seems reasonable to repeat the definition of computation tree here, which serves as a good interpretation of \emph{backtracking}.

\begin{definition}
Assume that $P$ is a BT problem and $\mathcal{A}$ is a BT algorithm for $P$, For any instance $I=(D_1,...,D_n), D_i \in \mathcal{D}_P$ we define the \emph{computation tree} $T_{\mathcal{A}}(I)$ as an oriented rooted tree in the following recursive way.
\begin{itemize}
\item Each node $v$ of depth $k$ in the tree is labelled by a tuple $(D_1^v,...,D_k^v,a_1^v,...,a_k^v)$.
\item The root node has the empty label.
\item For every node $v$ of depth $k<n$ with a label $(\overrightarrow{D}^v,\overrightarrow{a}^v)$, let $D_{k+1}^v$ be the data item in $I\setminus\{D_1^v,...,D_k^v\}$ that goes first in the list $r_{\mathcal{A}}^k(\overrightarrow{D}^v,\overrightarrow{a}^v)$. Assume that the output $c_{\mathcal{A}}^k(\overrightarrow{D}^v,D_{k+1}^v,\overrightarrow{a}^v)$ has the form $(c_1,...,c_d,\perp,c_{d+1},...)$, where $c_i \in H$. If $d=0$ then $v$ has no children. Otherwise it has $d$ child nodes $v_1,...,v_d$ that go from left to right and have labels\\
$(D_1^{v_i},...,D_{k+1}^{v_i},a_1^{v_i},...,a_{k+1}^{v_i})=(D_1^{v_i},...,D_{k+1}^{v_i},a_1^{v_i},...,a_k^{v_i},c_i)$ resp.
\end{itemize}
\end{definition}

Following is an example diagram of computation tree, nodes for data items, edges for choices. It's easy to see that it represents an adaptive BT algorithm, but not fully adaptive.

\begin{figure}[!h]
\begin{center}
\includegraphics{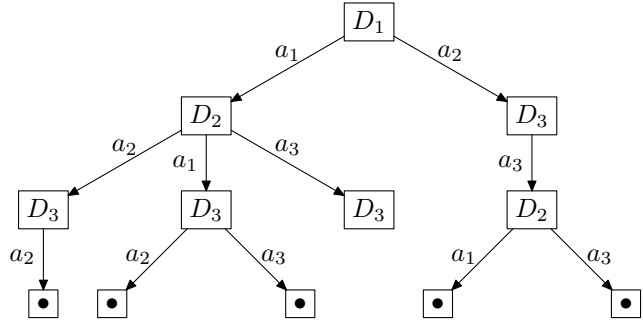}
\end{center}
\caption{A computation tree of some adaptive BT algorithm}
\label{Figure1}
\end{figure}

Clearly the width of a computation tree is a lower bound of the computational complexity of the corresponding BT algorithm, so we will consider the asymptotic lower bounds of the computation tree width in the following text.

\section{General lower bound strategy}

To understand the proof shown later, we give the basic idea to prove such lower bounds first. It's a game between Solver and Adversary. Initially, the Adversary presents to the algorithm(Solver) some finite set of possible input items, $P_0$, and the partial instance $PI_0$ is empty, $T_0$ is the set consisting of the null partial solution. The game consists of a series of phases. At any phase $i$, there is a set of possible data items $P_i$, a partial instance $PI_i$ and a set $T_i$ of partial solutions for $PI_i$. In phase $i$, $i \ge 1$, the Solver picks any data item $a \in P_{i-1}$, adds $a$ to obtain $PI_i=PI_{i-1}\cup\{a\}$, and chooses a set $T_i$ of partial solutions, each of which must extend a solution in $T_{i-1}$. The Adversary then removes $a$ and some further items to obtain $P_i$.

The strategy of Adversary will usually have the following form. The number of rounds, $n$ will be fixed in advance. The Adversary will choose some $i \le n$ such that, for many partial solutions $PS$ of $PI_i$, there is an extension of $PI_i$ to an instance $A \subset PI_i \cup P_i$ so that all valid/optimal/approximately optimal solutions to $A$ are extensions of $PS$. We'll call such a partial solution \emph{indispensable}, since if $PS \notin T_i$, the Adversary can set $P_i$ to $A \setminus PI_i$ so that the algorithm will never result in correct solution.

\section{Lower Bound of Knapsack}

First let us recall the definition of Knapsack problem.

\begin{description}
\item[Knapsack Problem] is defined by:

\begin{itemize}
\item Input: $n$ pairs of non-negative integers, $(x_1,p_1),...,(x_n,p_n)$ and a positive integer $N$, $x_i$ represents the weight of the $i$th item and $p_i$ represents the value of the $i$th item. $N$ is the volume of the knapsack.
\item Output: $S\subset\{1,...,n\}$, such that $\sum_{i \in S}{p_i}$ is maximized with respect to $\sum_{i \in S}{x_i} \le N$.
\end{itemize}

\item[Simple Knapsack Problem] is defined by:

\begin{itemize}
\item Input: $n$ non-negative integers $\{x_1,...,x_n\}$ and a positive integer $N$, $x_i$ is the weight and value of the $i$th item and $N$ is the volume of the knapsack.
\item Output: $S\subset\{1,...,n\}$, such that $\sum_{i \in S}{x_i}$ is maximized with respect to $\sum_{i \in S}{x_i} \le N$.
\end{itemize}
\end{description}

Both problems are NP complete, we will consider only the simple Knapsack problem, and denote a simple Knapsack problem with $n$ items and volume of $N$ with $(n,N)$.

M.Alekhnovich et. al \cite{1} have proved the following theorem:
\begin{theorem}
For simple Knapsack problem (n,N), the time complexity of any adaptive BT algorithms is at least
\begin{displaymath}
\binom{n/2}{n/4}=\Omega(2^{0.5n}/\sqrt{n})
\end{displaymath}
\end{theorem}

We will improve this lower bound using the same technique as the previous work, by optimizing related parameters, formally:
\begin{theorem}
For simple Knapsack problem (n,N), the time complexity of any adaptive BT algorithms is at least
\begin{displaymath}
\binom{n/2}{n/4}=\Omega(2^{(0.694-\epsilon)n}/\sqrt{n}),
\end{displaymath}
where $\epsilon$ is any small positive number.
\end{theorem}

\noindent\emph{Proof}: Consider positive numbers $\beta>\gamma$ satisfying $\beta+\gamma<1$, then $1-\beta>\gamma$, so we can choose another positive number $\alpha$ that satisfies $\alpha(1-\beta)>1$ and $\alpha\gamma<1$. These parameters will be fixed later to optimize the lower bound.

Let $N$ be some large integer, our initial set of items are integers in $I=(0,\alpha\cdot\frac{N}{n})$. Solver takes the first $\beta n$ items one by one, and following each one, Adversary applies the following rules to remove certain items from future consideration: remove all items that are the difference of the sums of two subsets already seen; also remove all items that complete any subset to exactly $N$ (ie all items with value $N-\sum_{x \in S}{x}$, where $S$ is a subset of the items considered so far).

These rules guarantee that at any point, no two subsets will generate the same sum, and that no subset will sum to $N$. Also notice that this eliminates at most $O(3^{\beta n})$ numbers. To know why, the difference of the sums of any two subsets can be represented as a weighted sum of the numbers seen so far, with three possible weights: 1, 0 and -1, so the number of distinct differences of any two subsets is at most $O(3^{\beta n})$; the number of distinct values that complete any subset to exactly $N$ is even less ($O(2^{\beta n})$), thus can be omitted here. So we will never exhaust the range from which Solver can pick the next item provided that $N>>3^{\beta n}$.

Call the set of numbers chosen so far $P$ and consider any subset $Q$ contained in $P$ of size $\gamma n$. Our goal is to show that $Q$ is indispensable; that is, we want to construct a set $R = R_Q$ of size $(1-\beta)n$, consisting of numbers in the feasible input with the following properties.

\begin{enumerate}
\item $\sum_{x \in Q \cup R}{x}=N$.
\item $P \cup R$ does not contain other subsets that sum to $N$.
\end{enumerate}

The above properties indeed imply that $Q$ is indispensable since obviously there is a unique solution with optimal value $N$ and, in order to get it, $Q$ is the subset that must be chosen among the elements of $P$. We thus get a lower bound on the width of the computation tree of any adaptive BT, which is the number of subsets of size $\gamma n$ in $P$; namely $\binom{\beta n}{\gamma n}$. Below is a diagram of such construction.

\begin{figure}[!h]
\begin{center}
\includegraphics{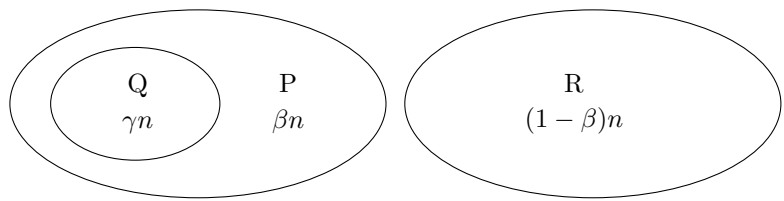}
\end{center}
\caption{Construction of an indispensable partial solution}
\end{figure}

For any $Q \subset P$, let $U=3^n$, $a=\frac{1}{(1-\beta)n}(N-\sum_{x \in Q}{x})$, $J=[a-U,a+U]$. We show that $J \subset I$. Note that numbers in $Q$ are less than $\alpha\cdot\frac{N}{n}$, so $a-U>\frac{1}{(1-\beta)n}(N-\gamma n\cdot\alpha\cdot\frac{N}{n})-3^n=\frac{1}{(1-\beta)n}(N-\gamma\alpha N)-3^n>0$, the last step follows from $\alpha\gamma<1$ and $N>>3^n$; and $a+U<\frac{1}{(1-\beta)n}\cdot N+3^n<\alpha\cdot\frac{N}{n}$, the last step follows from $\alpha(1-\beta)>1$ and $N>>3^n$.

To construct $R$, first choose $(1-\beta)n-2$ items in $J$. After each choice, we will remove some items from the remaining by the following rules. Let $S$ be the set of items currently in $P \cup R$.

\begin{enumerate}
\item For all $S_1,S_2 \subset S$, remove items of value $|\sum_{x \in S_1}{x}-\sum_{x \in S_2}{x}|$.
\item For all $S_1 \subset S$, remove items of value $N-\sum_{x \in S_1}{x}$.
\end{enumerate}

Since $U=3^n$, we can always avoid the points that need to be eliminated, and sum to a number $w$ so that $|w-a((1-\beta)n-2)| \le U$. This can be done by iteratively picking numbers bigger/smaller than $a$ according to whether they average to below/above $a$ currently.

To complete we need to pick two more items $b_1,b_2 \in I$ that sum to $v=(1-\beta)na-w$, and so that they are not the difference of sums of any two subsets of the $n-2$ items picked so far. Assume for simplicity that $v/2$ is an integer. Of the $U+1$ pairs $(v/2-i,v/2+i)$, where $i=1,2,...,U+1$, at least one pair $(b_1,b_2)$ will suffice.

(Note that, since $|v-2a| \le U$, the smallest number is $v/2-U-1 \ge a-\frac{3U}{2}-1>\frac{1}{(1-\beta)n}(N-\gamma\alpha N)-\frac{3U}{2}-1>0$, and the largest number is $v/2+U+1 \le a+\frac{3U}{2}+1<\frac{1}{(1-\beta)n}\cdot N+\frac{3U}{2}+1<\alpha\cdot\frac{N}{n}$, provided that $N$ is sufficiently large. So we will never go beyond the range of $I$.)

Now it has $\sum_{x \in Q \cup R}{x}=N$, the only thing left is to prove there does not exist another subset which also sums to $N$. Suppose for contradiction $S$ is such a subset, namely $\sum_{x \in S}{x}=N$, following are three cases:

\begin{enumerate}
\item Neither $b_1$ nor $b_2$ belong to $S$, contradictary to the second rule when picking the first $n-2$ items.
\item Both $b_1$ and $b_2$ belong to $S$, by comparing $S$ and $Q \cup R$, we get two subsets of the first $n-2$ items that sum equally, contradictary to the first rule.
\item Exactly one of $b_1$ and $b_2$ belong to $S$, by comparing $S$ and $Q \cup R$, we find that $b_1$ or $b_2$ is equal to the difference of the sums of some two subsets of the first $n-2$ items, contradiction.
\end{enumerate}

The rest of the proof is about maximizing the lower bound: $\binom{\beta n}{\gamma n}$, subject to $\beta>\gamma>0$ and $\beta+\gamma<1$.

According to Stirling formula, the target function can be simplified:
\begin{eqnarray}
\binom{\beta n}{\gamma n} & = & \frac{(\beta n)!}{(\gamma n)!((\beta-\gamma)n)!} \\ & \sim & \frac{\sqrt{2 \pi\beta n}(\frac{\beta n}{e})^{\beta n}}{\sqrt{2 \pi\gamma n}(\frac{\gamma n}{e})^{\gamma n}\cdot\sqrt{2 \pi(\beta-\gamma)n}(\frac{(\beta-\gamma)n}{e})^{(\beta-\gamma)n}} \\ & \sim & \frac{(\frac{\beta^\beta}{\gamma^\gamma\cdot(\beta-\gamma)^{\beta-\gamma}})^n}{\sqrt{n}}
\end{eqnarray}

By applying logorithm transform, our goal is equivalent to maximize:
\begin{displaymath}
f(\beta,\gamma)=\beta ln\beta-\gamma ln\gamma-(\beta-\gamma)ln(\beta-\gamma),
\end{displaymath}
subject to $\beta>\gamma>0$ and $\beta+\gamma \le 1$.

Note that for convenience, we allow $\beta+\gamma=1$ here, the case will be handled later. Take partial differential on $\beta$, which is:
\begin{displaymath}
\frac{\partial f(\beta,\gamma)}{\partial \beta}=ln\beta-ln(\beta-\gamma)>0
\end{displaymath}

So $f(\beta,\gamma)$ is strictly increasing respect to $\beta$, since $\beta \le 1-\gamma$, by setting $\beta$ to $1-\gamma$, we only need to maximize:
\begin{displaymath}
g(\gamma)=f(1-\gamma,\gamma)=(1-\gamma)ln(1-\gamma)-\gamma ln\gamma-(1-2\gamma)ln(1-2\gamma)
\end{displaymath}

Whose differential is:
\begin{displaymath}
g'(\gamma)=2ln(1-2\gamma)-ln\gamma-ln(1-\gamma)=ln\frac{(1-2\gamma)^2}{\gamma(1-\gamma)}
\end{displaymath}

The only root of the above function that lies in $(0,\frac{1}{2})$ is $\frac{5-\sqrt{5}}{10}$, which maximizes $g$, thus $f$. So $\beta$ can be arbitrarily close to $1-\gamma=\frac{5+\sqrt{5}}{10}$, and the maximum value of $\frac{\beta^\beta}{\gamma^\gamma\cdot(\beta-\gamma)^{\beta-\gamma}}$ is $1.618-\epsilon$ by substitution, where $\epsilon$ is any small positive number. Consequently, the optimal lower bound $\binom{\beta n}{\gamma n}$ is:
\begin{displaymath}
(1.618-\epsilon)^n/\sqrt{n} \doteq 2^{(0.694-\epsilon)n}/\sqrt{n}
\end{displaymath}

\section{Discussion and Future Work}

Backtracking is an important algorithmic scheme which is pervasive in solving hard problems such as Propositional Satisfiability Problem (SAT) and Constraint Satisfaction Problem (CSP). Proving lower bounds for these problems under BT model will both deepen our understanding of the structure and properties of these hard problems and guide our designing of algorithms to solve these problems. An exponential time lower bound for SAT under fully adaptive BT model has already appeared in \cite{1}. Recently, Xu and Li have proven an exponential lower bound for a class of random CSPs (Model RB) under tree-like resolution \cite{5}. Since tree-like resolution has a close tie with DPLL, a famous backtracking search strategy for SAT with corresponding exponential time lower bound in \cite{4}, it is natural to investigate the time complexity of some class of random CSPs (RB model) under BT model, and we expect that this future work may produce the first exponential time lower bound for some class of random CSPs under BT model.

\section{Acknowledgements}
We thank Professor Ke Xu for joining our discussion and providing helpful comments.


\begin{thebibliography}{00}

\bibitem{1} A.Borodin, A.Magen, J.Buresh-Oppenheim, M.Alekhnovich, R.Impagliazzo and T.Pitassi: Toward a Model for Backtracking and Dynamic Programming, Proceedings of 20th Annual IEEE Conference on Computational Complexity (2005) 308-322
\bibitem{2} Xin Li, Tian Liu, Han Peng, Liyan Qian, Hongtao Sun, Jin Xu, Ke Xu and Jiaqi Zhu: Improved Exponential Time Lower Bound of Knapsack Problem under BT Model, Proceedings of Fourth Annual Conference on Theory and Applications of Models of Computation (2007) 624-631
\bibitem{3} P.Pudlak: Proofs as games. American Math. Monthly, 23 (2000) 541-550
\bibitem{4} D.Itsykson, E.Hirsch and M.Alekhnovich: Exponential Lower Bounds for the Running Time of DPLL Algorithms on Satisfiable Formulas, Proceedings of 31st International Colloquium on Automata, Languages and Programming (2004) 84-96
\bibitem{5} Ke Xu, Wei Li: Many Hard Examples in Exact Phase Transitions, Theoretical Computer Science, 355 (2006) 291-302
\bibitem{6} A.Borodin, C.Rackoff and M.Nielson: (Incremental) Priority Algorithms, Algorithmica, 37 (2003) 295-326
\bibitem{7} G.Woeginger: When Does a Dynamic Programming Formulation Guarantee the Existence of a Fully Polynomial Time Approximation Scheme (FPTAS)? INFORMS Journal on Computing, 12 (200) 57-75
\bibitem{8} A.Borodin: Further Reflections on a Theory for Basic Algorithms, Proceedings of Second Annual Conference on Algorithmic Aspects in Information and Management (2006) 1-9
\end{thebibliography}
\end{document}